\documentclass{PoS}
\rightline{\sffamily UTCCS-P-48, UTHEP-576}\vspace*{-10mm}
\title{Making use of the International Lattice Data Grid }

\ShortTitle{Making use of the International Lattice Data Grid }

\author{\speaker{Tomoteru Yoshie}\\%
        Center for Computational Sciences, University of Tsukuba,
        Tsukuba 305-8577, Japan\\
        E-mail: \email{yoshie@ccs.tsukuba.ac.jp}}

\abstract{
The International Lattice Data Grid (ILDG) continues stable operation
for about one year and has accumulated a lot of valuable configurations.
After a brief review of the ILDG system, we highlight large
physics projects, whose configurations are already available on the
grid or will be open to the public in the near future. 
With such information, one can make better use of the ILDG.
Statistics about the ILDG is also reported.}          

\FullConference{The XXVI International Symposium on Lattice Field Theory \\
		 July 14 - 19, 2008\\
		 Williamsburg, Virginia, USA}

\begin{document}

\section{Introduction}
Six years ago at Lattice 2002 the construction of the International Lattice
Data Grid was proposed~\cite{ref:ILDG2002}.
After extensive work~\cite{ref:ILDG2003467,ref:MDWG2007,ref:MWWG2007} 
by many people, the first stage system construction completed last year. 
The ILDG is already used to open data to the public
and to share data within collaboration.
With this situation in mind, we'd like to invite new users
to the ILDG. 

For this purpose, we give a brief system overview in 
sec.~\ref{sec:overview}, describe how to use data on the grid in
sec.~\ref{sec:use}, and highlight major ensembles on the grid in
sec.~\ref{sec:data}, so that one can start to find interesting
ensembles.
We also give statistics about the ILDG in sec.~\ref{sec:stat}.

\section{System overview}\label{sec:overview}
The ILDG bears two aspects, metadata and middleware. 
Work on the ILDG has been carried out by the corresponding
working groups. Current members of the working groups are
\begin{itemize}
\item Metadata working group (MDWG):\ \ 
P.~Coddington (Adelaide), T.~Yoshie (Tsukuba), D.~Pleiter (DESY),
G.~Andronico (INFN), C.~Maynard (Edinburgh), C.~DeTar (Utah),
J.~Simone (FNAL), R.~Edwards, B.~Joo (JLAB)
\item Middleware working group (MWWG):\ \  
P.~Coddington, S.~Zhang (Adelade), T.~Amagasa, N.~Ishii. 
O.~Tatebe. M.~Sato (Tsukuba), D.~Melkumyan, D.~Pleiter (DESY), 
G.~Beckett, R.~Ostrowski (Edinburgh), J.~Simone (FNAL), 
B.~Joo, C.~Watson (JLAB)
\end{itemize}
In addition, the ILDG board, which consists of one representative from
each country, supervises two working groups and decides strategic
issues. Current members are
\begin{itemize}
\item ILDG board:\ \ 
R.~Brower (USA), K.~Jansen (Germany), R.~Kenway (UK, chair this year),
D.~Leinweber (Australia), O.~Pene (France), F.~Di~Renzo (Italy), 
A.~Ukawa (Japan)
\end{itemize}

\begin{figure}[t]
\begin{center}
\includegraphics[scale=0.22]{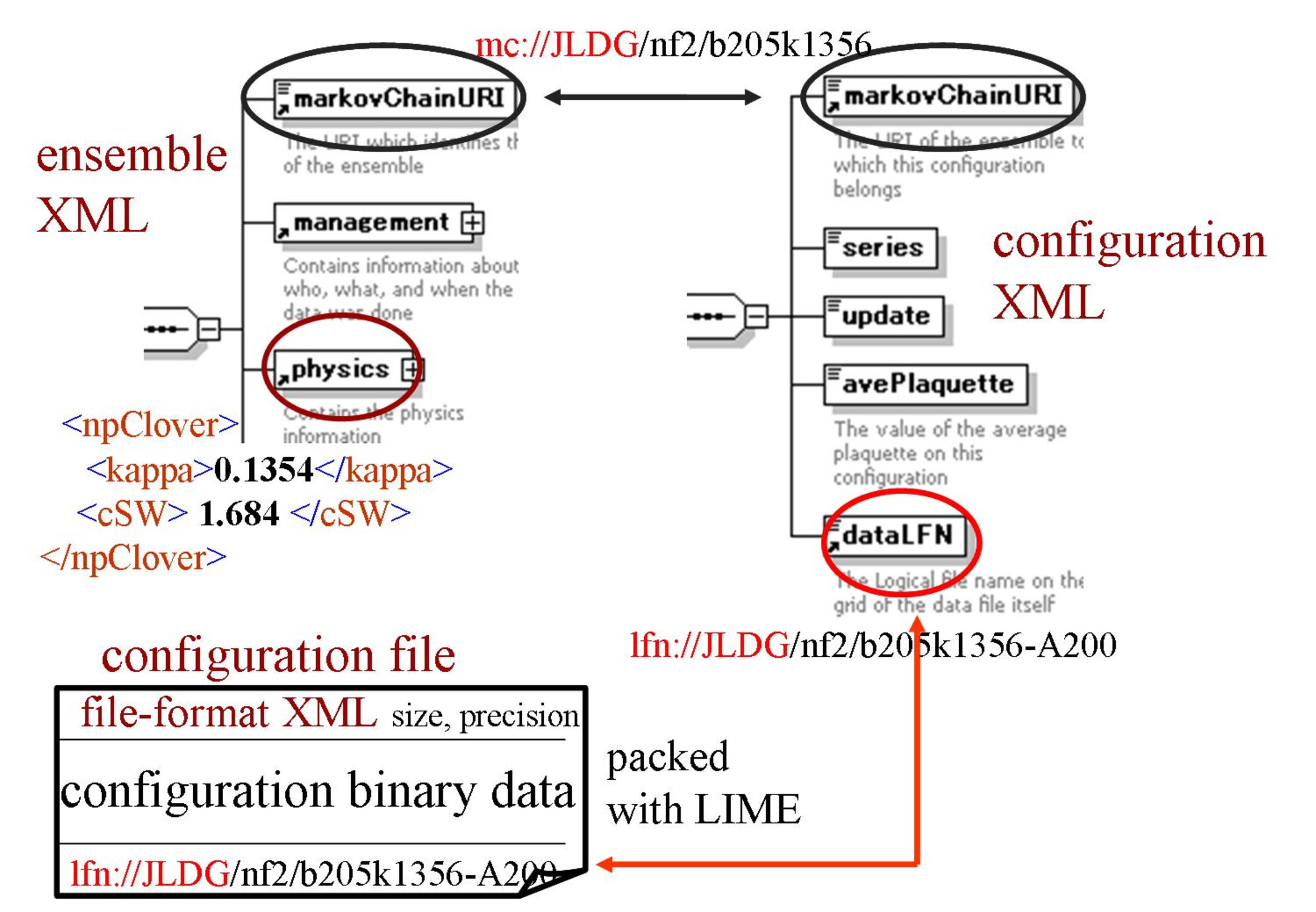}
\caption{Metadata and data components and how they are linked.}
\label{fig:metadata}
\end{center}
\end{figure}

Figure \ref{fig:metadata} sketches data and metadata components of the ILDG.
Properties (metadata) of an ensemble such as physics parameters of the
simulation and those of a configuration such as the trajectory number
are marked up with the QCDml~\cite{ref:MDWG2007,ref:MDWGweb},
an XML based markup language. 
The ensemble XML and the configuration XML are linked by a unique
name of the ensemble called ``markovChainURI''. 
The configuration XML and a configuration file are
linked by a unique name of the configuration called 
``dataLFN'' (data Logical File Name).

\begin{figure}[h]
\begin{center}
\includegraphics[scale=0.20]{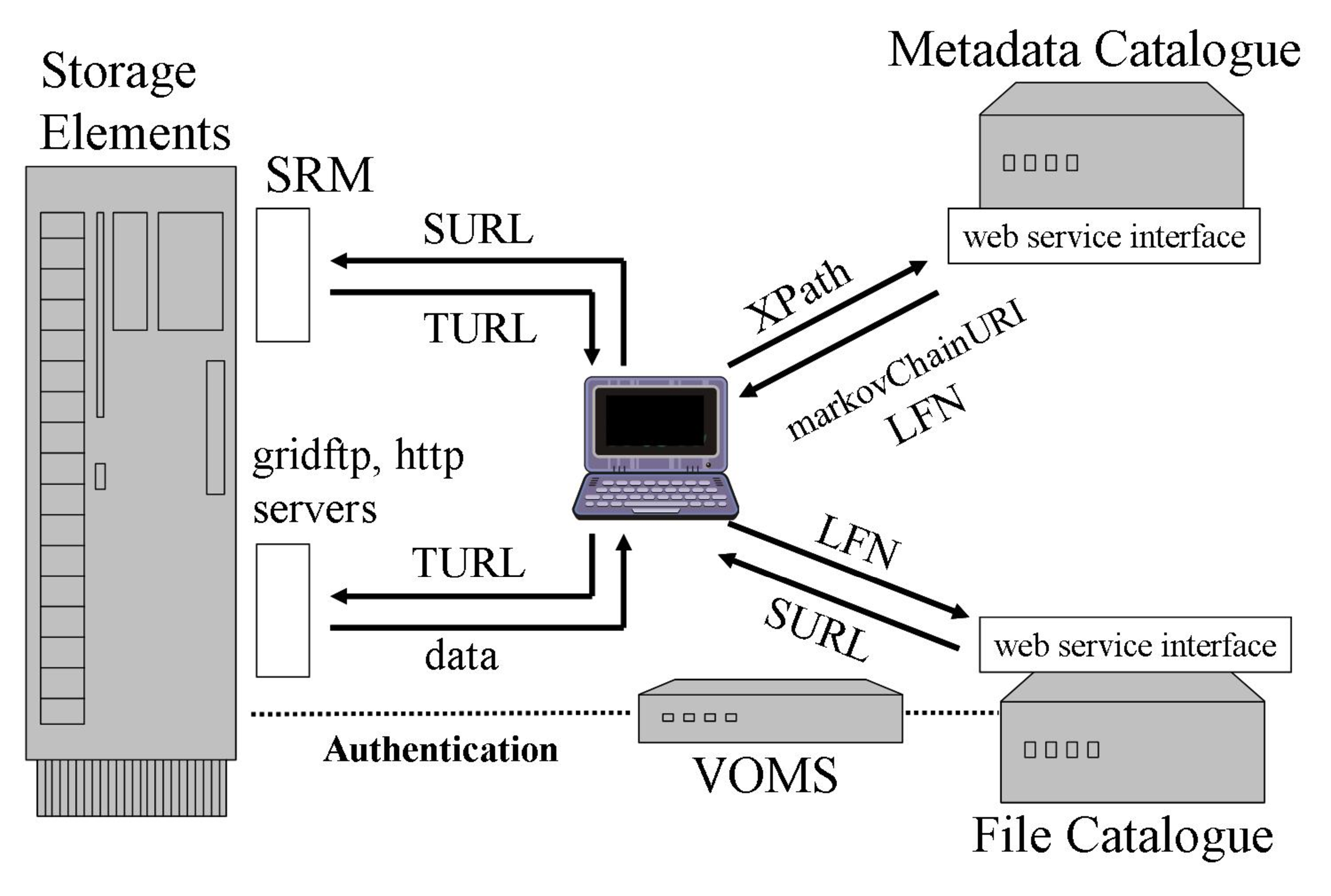}
\caption{Middleware components and data-metadata flow.}
\label{fig:middleware}
\end{center}
\end{figure}

Figure \ref{fig:middleware} summarizes middleware components of the
ILDG and data-metadata flow~\cite{ref:MWWG2007}. 
Configurations are stored in ``Storage Elements'', while metadata are
stored in the ``Metadata Catalogue''. 
Queries on ensemble and configuration metadata are made to the
Metadata Catalogue. 
The ``File Catalogue'' maps the dataLFN to one or more locations for the
file, or URLs of the file.  
After you get an URL, you can download the configuration file from
one of storage elements. 
Authentication is made before you download. 
The VOMS (Virtual Organization Membership Service) server maintains
the ILDG virtual organization (VO). 
The ILDG consists of five regional grids (RG), CSSM for Australia, 
JLDG for Japan, LDG (LatFor Data Grid) for continental Europe, 
UKQCD for UK, and USQCD for US.
Implementations of storage elements and catalogue services are 
different grid by grid, but RGs are inter-operable with a common
interface. 

\section{Using data on ILDG}\label{sec:use}
Although the ILDG system is a little bit complicated, users don't have
to remember details, because easy-to-use tools are developed and
provided by the MWWG.  

A procedure to use data on the ILDG is as follows.
\begin{enumerate}
\item Join the ILDG VO. 
To do this, one obtains a grid certificate from a CA (Certificate
Authority) trusted by the IGTF (International Grid Trust 
Federation~\cite{ref:IGTF}), and visits the ILDG VOMRS
(VO Member Registration Service~\cite{ref:VOMRS}) to register. 
A representative of your regional grid will approve your registration request.
The certificate is necessary when you download files.
\item Find interesting ensembles. You can use portals or tools
provided by RGs, which can be accessed from the ILDG official web
page~\cite{ref:ILDGweb}. Some details will be given below.
\item Check access policy before you use data. 
Data on the ILDG are either public or can be used
after negotiation with the collaboration. 
The best way to know the policy is to contact the collaboration. 
\item Download configurations. One can use a standard command line 
tool {\bf ildg-get} provided by the MWWG. 
Note that some RGs support other methods to access data, such as
ltools (LDG), DiGS tools (UKQCD) and uberftp (JLDG).
\item Do research and write a paper. Please acknowledge the
collaboration by citing papers specified by the collaboration and  
the ILDG web page ``http://www.lqcd.org/ildg/''. 
\end{enumerate}
In addition to RG portals, a web page~\cite{ref:tutorial} prepared
for the ILDG tutorial session of this conference, which is organized
by C.~Urbach and C.~Allton, provides you with a good starting point.

\begin{figure}[t]
\begin{center}
\includegraphics[scale=0.31]{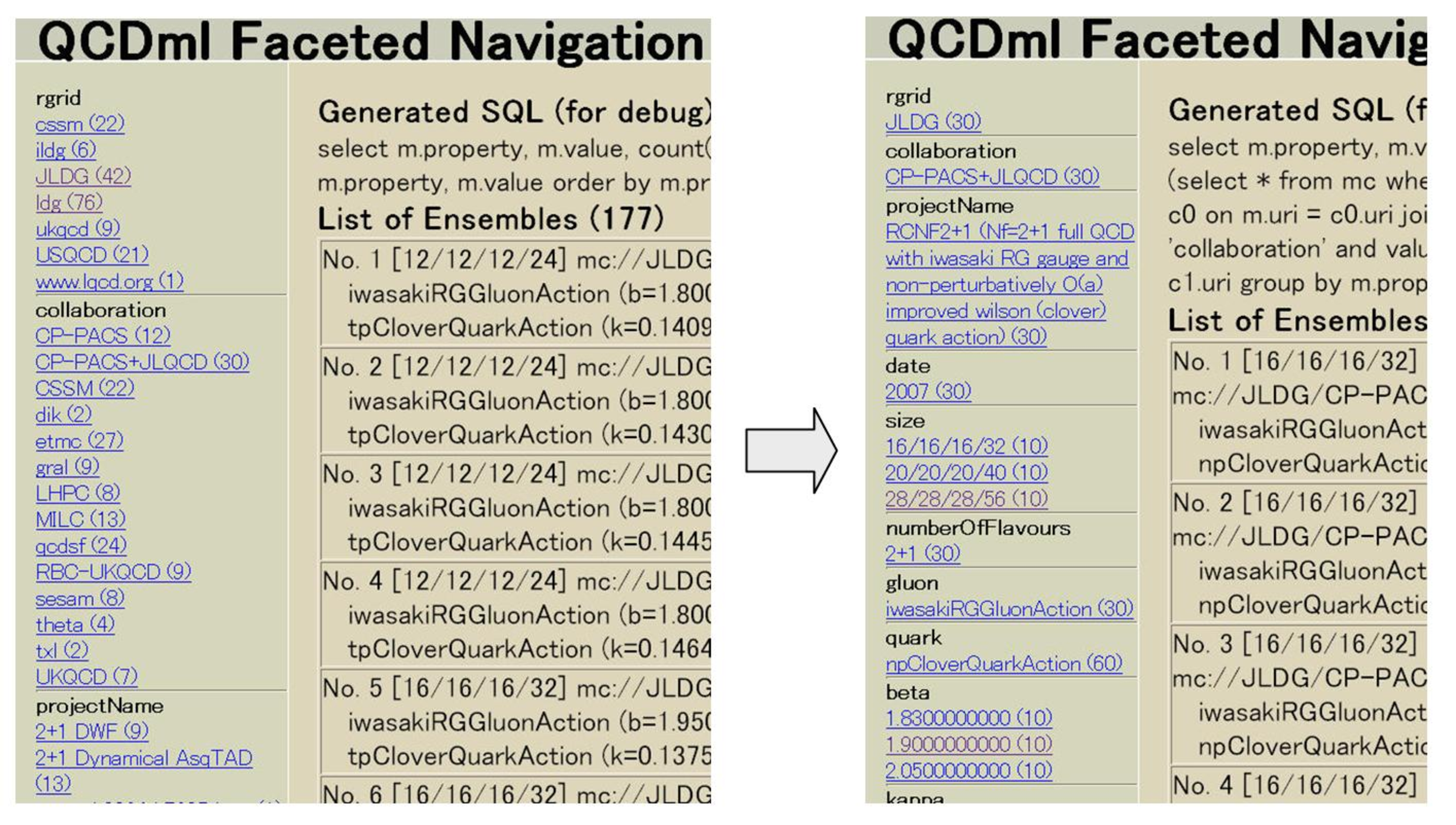}
\caption{A screen shot of faceted navigation, taken from the new JLDG portal.}
\label{fig:JLDGportal}
\end{center}
\end{figure}

In order to help users to find ensembles, RGs provide variety of 
data browsers and tools. 
The USQCD portal~\cite{ref:USQCDportal} 
and the LDG portal~\cite{ref:LDGportal}
show you a list of ensembles and details of each ensemble.  
The CSSM portal~\cite{ref:CSSMportal} enables you to search ensembles
by specifying actions and other physics parameters. 
The UKQCD ildg-browser~\cite{ref:UKQCDportal} supports semantic search
based on XML.  
A new JLDG portal, which is still under development
and will appear soon in \cite{ref:JLDGportal}, will support
narrowing search by faceted navigation. 
Facets are categories of XML documents,
such as names of regional grid, collaboration, project etc., and
physics parameters. 
You will be able to specify any items in any order for narrowing
candidates.
See a screen shot in Fig.~\ref{fig:JLDGportal}.
You can use portals without joining the ILDG VO. 
Please visit and try all portals freely and find your favorite one.

\section{Ensembles on the grid}\label{sec:data}
This section summarizes major ensembles which are already 
available or will appear in the near future. 
We have asked several collaborations to list interesting ensembles,
and have compiled replies. 
Therefore, what is shown below is not a complete list and may be
biased due to our queries.

In the following tables, we use abbreviations for the action field,
{\it Sym}: Symanzik, {\it LW}: Luescher-Weisz, {\it np}:
non-perturbatively $O(a)$ improved, {\it tp}: tadpole improved, and
{\it TM}: twisted mass. The number of configurations is approximate and 
the status field indicates status of
ensembles by key words {\it public}: publicly available, 
{\it negotiable}: available but one can use them after negotiation
with the collaboration, {\it prod.}: production run is on-going, and
{\it prep.}: configurations are in preparation and will be available soon.   

\subsection{CSSM}
Table \ref{tab:CSSM} lists ensembles on the CSSM grid.
The CSSM collaboration has started to accumulate $N_f=2$ 
configurations generated with the FLIC fermion action. 
We hear that they continue parameter tuning for a while and plan
to quantify the advantages of the FLIC action for dynamical quark 
simulations. 

The CSSM grid has a 7 TB disk and 20 TB tape system for storage elements.
Catalogue services are operated at CSSM.

\begin{table}
\begin{center}
\begin{tabular}{c|c|c|c|c|c|c|l}
 \hline\hline 
 $N_f$ & action & collab &  $a$ (fm) & lattice & $m_\pi$ (MeV) & conf & status \\
 \hline
  0 & /tpLW,DBW2 & CSSM & & & & 1500 & public \\
 \hline
  2 & FLIC/tpLW  & CSSM & 0.096 & $16^3\times 32$ & 820 & 50 & public\\
 \hline
  2 & FLIC/tpLW  & CSSM & 0.125 & $20^3\times 40$ & $>300$ & & prod. \\
 \hline\hline
\end{tabular}
\caption{Major ensembles on the CSSM grid.}
\label{tab:CSSM}
\end{center}
\end{table}

\subsection{JLDG}
Table \ref{tab:JLDG} shows ensembles on the JLDG.
In addition to the CP-PACS $N_f=2$ and the CP-PACS+JLQCD $N_f=2+1$ 
ensembles already public to the world, the $N_f=2$ and $N_f=2+1$
overlap quark ensembles and the $N_f=2+1$ npClover ensembles
with very light quark masses will be publicly released by the JLQCD
collaboration and the PACS-CS collaboration, respectively.

Storage elements, with 35 TB disk space in total, are distributed over
six sites in Japan, Tsukuba, KEK, Kyoto, Osaka, Hiroshima and
Kanazawa. Catalogue services are operated at CCS, Tsukuba. 

\begin{table}
\begin{center}
\begin{tabular}{c|c|c|c|c|c|c|l}
 \hline\hline 
  $N_f$ & action & collab &  $a$ (fm) & lattice & $m_\pi$ (MeV) & conf & status \\
 \hline 
  2 & tpClover & CP-PACS & 0.22 & $12^3\times 24$ & 1060$-$490 &
 1000$\times 4$ & public \\
 \cline{4-7}
    &   /Iwasaki       &         & 0.16 & $16^3\times 32$ & 1270$-$540 &
 1000$\times 4$ & \\
 \cline{4-7}
    &                &         & 0.11 & $24^3\times 48$ & 1160$-$540 &  
  800$\times 4$ &   \\
 \hline
 2 & npClover/Plaq. & JLQCD & 0.09 & $20^3\times 48$ & 1370$-$600 & 
  1200$\times 5$ &  prep.(*1)\\
 \hline
 2+1 & npClover& CP-PACS & 0.12 & $16^3\times 32$ & 1200$-$620
 & 800$\times 5 \times 2$ & public \\
 \cline{4-7}
     &   /Iwasaki     &  +JLQCD  & 0.10 & $20^3\times 40$ & 1100$-$650
 & 800$\times 5 \times 2$ & \\
 \cline{4-7}
     &                &         & 0.07 & $28^3\times 56$ & 1030$-$630
 & 600$\times 5 \times 2$ & \\
 \hline
 2 & Overlap/Iwasaki & JLQCD & 0.12 & $16^3\times 32$ & 750$-$290 &
 500$\times 6$ & prep.(*1) \\
 \hline
 2+1 & Overlap/Iwasaki & JLQCD & 0.11 & $16^3\times 48$ & 800$-$310 &
 500$\times 5 \times 2$ & (*2) \\
 \hline
 2+1 & npClover & PACS-CS & 0.09 & $32^3\times 64$ & 702$-$156 &
 400$\times 4$ & (*3) \\
     & /Iwasaki   &         &      &                 &           &
 800$\times 2$ \\
\hline\hline
\end{tabular}
\caption{Major ensembles on the JLDG. (*1: will be public,
*2: available date not decided yet, *3 will be public 6 months after a
spectrum paper is submitted.)} 
\label{tab:JLDG}
\end{center}
\end{table}

\subsection{LDG}

For the LatFor data grid, ensembles generated by two major
contributors are listed in table \ref{tab:LDG}. 
The ETM collaboration has carried out $N_f=2$ simulations with
the Wilson twisted mass quark action for three lattice spacings and
for a wide range of quark masses. 
All of these configurations are already put on the grid and can be
used after negotiation with the collaboration. 
We hear that they will become publicly available probably by the end
of this year. 
The collaboration has also started $N_f=2+1+1$ simulations. 
$N_f=2$ configurations from the QCDSF collaboration, also
cover wide ranges of lattice spacings and quark masses, need
negotiation. 
$N_f=2+1$ simulations with the SLiNC quark action is on-going. 
The LDG contains data from other collaborations such as SESAM,
T$\chi$L, GRAL, DIK, and Theta.
We hear that the ALPHA collaboration has no plan to submit data, 
and the Bern-Marseilles-Wuppertal collaboration has not made a
decision about their plans for the ILDG.

The LatFor data grid is for continental Europe. Storage elements are
distributed over 3 sites in Germany, DESY (Hamburg+Zeuthen), JSC
(J\"{u}lich), ZIB (Berlin), CC-IN2P3 in Lyon, France and INFN Parma in Italy.
Storage elements have tape back-end without fixed storage quota. 
The RG operates catalogue services at DESY.

\begin{table}
\begin{center}
\begin{tabular}{c|c|c|c|c|c|c|l}
 \hline\hline 
 $N_f$ & action & collab &  $a$ (fm) & lattice & $m_\pi$ (MeV) & conf & status \\
 \hline
 2 & wilson-TM & ETM  & 0.100 & $20^3\times 48$ & 700$-$300 &
 2000$\times 4$ & negotiable \\ 
 &/Sym&&& $24^3\times 48$ &  700$-$300 & 2000$\times 5$ & (*1) \\ 
 \cline{4-7}
 &&& 0.085 & $24^3\times 48$ & 700$-$300 & 2500$\times 5$ & \\
 &&&       & $32^3\times 64$ & 300$-$250 & 2500$\times 2$ & \\
 \cline{4-7}
 &&& 0.066 & $20^3\times 48$ & 400$-$280 & 3000$\times 2$ & \\
 &&&       & $24^3\times 48$ & 350       & 3000           & \\
 &&&       & $32^3\times 48$ & 700$-$280 & 2500$\times 4$ & \\
 \hline
2+1+1 & wilson-TM & ETM & 0.090 & $24^3\times 48$ & 700$-$300 & O(1000) & prod. \\
 &/Sym &&&&& \\    
 \hline
2 & npClover & QCDSF & 0.11 & $16^3\times 32$ & 1200$-$250 & 19 ensembles & negotiable \\
 & /wilson && $-$0.07 & $-40^3\times 64$ & & $\approx$ 20000 (*2) & \\
\hline
2+1 & SLiNC & QCDSF & 0.08 & $48^3\times 64$ & 500$-$200 & & prod. \\
    &/Sym &&&&& \\
 \hline\hline
\end{tabular}
\caption{Major ensembles on the LDG.
(*1: become publicly available probably by the end of 2008,
 *2: based on Metadata Catalogue)}
\label{tab:LDG}
\end{center}
\end{table}

\subsection{UKQCD}
The UKQCD grid (see table \ref{tab:UKQCD}) contains $N_f=2+1$ domain
wall ensembles generated by the joint collaboration of the UKQCD and
the RBC. $16^3$ lattices were publicly available before this conference. 
The collaboration has publicly released configurations on $24^3$
lattices this August.
Simulations on finer lattices are on-going. 
The grid has also archived the UKQCD asqtad lattices.
Configurations on the coarser lattice are publicly available, while 
those on the finer lattice are negotiable.

The UKQCD grid consists of seven sites in UK,
Edinburgh, ACF (U. of Edinburgh), Glasgow, Liverpool, Swansea, 
RAL Didcot and Southampton.
The grid maintains 80 TB disk space (as of March 2007) for storage elements.
Catalogue services are operated at EPCC, Edinburgh.

\begin{table}
\begin{center}
\begin{tabular}{c|c|c|c|c|c|c|l}
 \hline\hline 
 $N_f$ & action & collab &  $a$ (fm) & lattice & $m_\pi$ (MeV) & conf & status \\
 \hline
2+1 & Domain Wall & UKQCD & 0.12 & $16^3\times 32 \times 16$ & 630 & 1517 & public \\
    & /Iwasaki & /RBC & & & 530 & 810  \\
    &          & & & & 400 & 832  \\
 \cline{4-8}
 &&& 0.12 & $24^3\times 64\times 16$ & 670$-$330 & 800$\times 4$ & public(*1) \\
 \cline{4-8}
 &&& 0.08 & $32^3\times 64\times 16$ & 400$-$280 & & prod. \\
 &&&      & $48^3\times 64\times 16$ & $\approx$ 220 & &  \\
 \hline
2+1 & asqtad/tpSym & UKQCD & 0.12 & $24^3\times 64$ & 290 & 5081 & public \\
 \cline{4-8}
    & & & 0.09 & $32^3\times 64$ & 360 & 700 & negotiable \\
 \hline\hline
\end{tabular}
\caption{Major ensembles on the UKQCD grid. 
(*1: public release in this August)}
\label{tab:UKQCD}
\end{center}
\end{table}

\subsection{USQCD}
Table \ref{tab:USQCD} summarizes ensembles in the USQCD grid.
The MILC collaboration has been generating an extensive set of 
$N_f=2+1$ ensembles using the asqtad quark action. A remarkable point is
that the collaboration makes all data public as soon as they are
created.  
Currently, they are generating data on large lattices at 
0.12 and 0.09 fm, and on much finer lattices. They will be 
available on the grid or the NERSC Gauge Connection web 
site~\cite{ref:GCweb}. 
The LHP collaboration is generating anisotropic lattices. 
$N_f=2$ data are already publicly available, and $N_f=2+1$
data are coming soon. 

Storage elements of this grid are 
operated at Fermilab as a part of huge resources, while catalogue 
services are operated at JLAB.

\begin{table}
\begin{center}
\begin{tabular}{c|c|c|c|c|c|c|l}
 \hline\hline 
 $N_f$ & action & collab &  $a$ (fm) & lattice & $m_\pi$ (MeV) & conf & status \\
 \hline
2+1 & asqtad & MILC & 0.15 & $(16-20)^3\times 48$ & 711$-$235 & 600$\times 4$ & public \\
 \cline{4-8}
    &/tpLW&& 0.12 & $(20-24)^3\times 64$ & 500$-$260 & 1700$\times 4$ & public \\
 \cline{5-8}
    &&&      & $32^3\times 64$ & $\approx$ 260 &  & prod. \\
 \cline{4-8}
    &&& 0.09 & $(28-40)^3\times 96$ & 480$-$240 & 1100$\times 6$ & public, prod. \\
 \cline{5-8}
    &&&      & $40^3\times 96$ & $\approx$ 240 &  & prod. \\
 \cline{4-8}
    &&& 0.06 & $(48-64)^3\times 144$ & 430$-$220 & 600$\times 4$ & public, prod. \\
 \cline{4-8}
    &&& 0.045 & $64^3\times 192$ & TBD. & 300 & public, prod. \\
 \hline
2 & aniso. & LHP & 0.11 & $16^3\times 64$ & 600 & 861 & public \\
 \cline{5-7}
  & wilson & & & $24^3\times 64$ & 600 & 871 & \\  
  & /wilson& & &                 & 440 & 1535 & \\  
 \hline
2+1 & (*1) & LHP & 0.12 & $24^3\times 128$ & 330 & 2000 & prep. \\
 \hline\hline
\end{tabular}
\caption{Major ensembles on the USQCD grid. 
(*1: aniso. clover/ tadpole improved Symanzik with no rectangle loops in
temporal direction.)} 
\label{tab:USQCD}
\end{center}
\end{table}

\section{Statistics}\label{sec:stat}

\begin{table}
\begin{center}
\begin{tabular}{l||r|r|r|r}
 \hline\hline 
 RG & \#VO member & \#ensemble & \#config. & data size (TB)\\
 \hline
 CSSM &   7 &  22 &   1.7K &  0.1 \\
 JLDG &   4 &  42 &  29.2K &  7.5 \\
 LDG  &  49 &  82 & 142.6K & 28.0 \\
 UKQCD&  26 &  16 &   8.8K &  2.9 \\
 USQCD&   7 &  21 &  10.9K &  2.7 \\
 \hline
 total&  93 & 183 & 193 K  &  41 \\
 \hline\hline
\end{tabular}
\caption{Statistics about the ILDG as of June 26, 2008. 
Data are taken from VOMRS and Metadata Catalogues.
Data size does not include file replica.}
\label{tab:statistics}
\end{center}
\end{table}

In order to see how the ILDG is utilized, statistics about the ILDG
is summarized in table \ref{tab:statistics}.

We have 93 ILDG VO members in total. Because the LDG and the UKQCD
grid have many users, we suppose that they use the ILDG as 
their primary storage infrastructure. 
The CSSM grid and the USQCD grid have genuine users. 
The JLDG has only admin users. We hope may Japanese users, who
still use an old system, will move to the ILDG.

Number of ensembles stored in the ILDG increases almost linearly since
January 2006 and have reached 183. 
We currently have $\approx$ 190 K configurations with total size 
of $\approx$ 40 TB.

\section{Conclusions and future work}
The ILDG continues stable operation and has already accumulated a lot
of valuable configurations.  
Usability of the ILDG is improved significantly.
The ILDG is becoming an important research 
infrastructure in this community.
We hope that many more users join the ILDG and make better use of
archived data for physics researches.

In this report, we have described how to use data on the ILDG.
Submitting data to the ILDG is a somewhat complicated procedure.
Working group members think that making the procedure easy is
an important future direction. 
Two working groups are also discussing extensions and improvements 
of the system, such as quark propagator sharing (MDWG) and
replication of data among regional grids (MWWG). 

I am grateful to all members of the ILDG working groups and 
the ILDG board,
in particular to G.~Beckett, C.~DeTar, and D.~Pleiter for helpful
suggestions on the manuscript. 
I also thank colleagues who provided us with ensemble information on
each regional grid.
A part of this work is supported by the Grant-in-Aid of the Ministry
of Education (No. 18104005 
) of the Japanese Government.

\end{document}